\documentclass[manuscript]{acmart}
\usepackage{makecell}
\usepackage{booktabs}
\usepackage{longtable}
\usepackage{tabularx}

\copyrightyear{2023}
\acmYear{2023}
\setcopyright{acmlicensed}\acmConference[CHI '23]{Proceedings of the 2023 CHI Conference on Human Factors in Computing Systems}{April 23--28, 2023}{Hamburg, Germany}
\acmBooktitle{Proceedings of the 2023 CHI Conference on Human Factors in Computing Systems (CHI '23), April 23--28, 2023, Hamburg, Germany}
\acmPrice{15.00}
\acmDOI{10.1145/3544548.3581359}
\acmISBN{978-1-4503-9421-5/23/04}

\begin{document}

\title[TactIcons: 3D Printed Map Icons for BLV]{TactIcons: Designing 3D Printed Map Icons for People who are Blind or have Low Vision}
\author{Leona Holloway}
\orcid{0000-0001-9200-5164}
\affiliation{
    \institution{Monash University}
    \city{Melbourne}
    \country{Australia}}
\email{leona.holloway@monash.edu}
\author{Matthew Butler}
\orcid{0000-0002-7950-5495}
\affiliation{
    \institution{Monash University}
    \city{Melbourne}
    \country{Australia}}
\email{matthew.butler@monash.edu}
\author{Kim Marriott}
\orcid{0000-0002-9813-0377}
\affiliation{
    \institution{Monash University}
    \city{Melbourne}
    \country{Australia}}
\email{kim.marriott@monash.edu}

\begin{abstract}
Visual icons provide immediate recognition of features on print maps but do not translate well for touch reading by people who are blind or have low vision due to the low fidelity of tactile perception. 
We explored 3D printed icons as an equivalent to visual icons for tactile maps addressing these problems.
We designed over 200 tactile icons (TactIcons) for street and park maps. These were touch tested by blind and sighted people, resulting in a corpus of 33 icons that can be recognised instantly and a further 34 icons that are easily learned. Importantly, this work has informed the creation of detailed guidelines for the design of TactIcons and a practical methodology for touch testing new TactIcons. 
It is hoped that this work will contribute to the creation of more inclusive, user-friendly tactile maps for people who are blind or have low vision. 
\end{abstract}

  \begin{teaserfigure}
    \includegraphics[width=\textwidth]{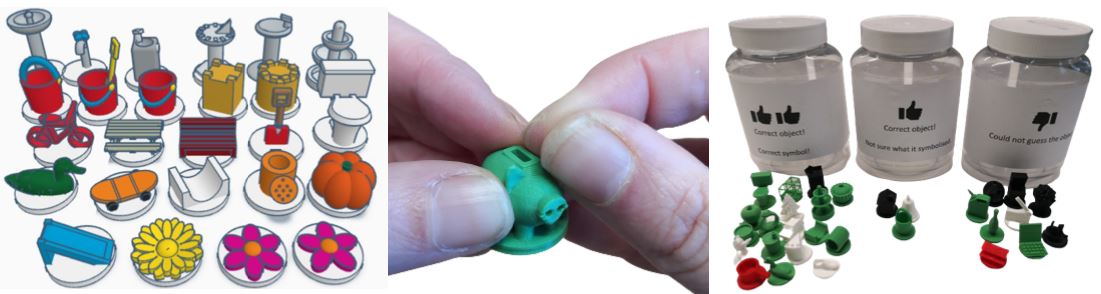}
    \caption{Design and touch testing of 3D printed tactile icons for maps of parks, playgrounds and shops.}
    \Description{Three images: 1. 3D models for tactile icons on circular bases for use on maps of parks and playgrounds. 2. Hands touching a small 3D printed piggy bank (a prototype icon to represent a bank). 3. 3D printed icons sorted in front of jars labelled with two thumbs up, one thumb up, or a thumbs down symbol.}
    \label{fig:teaser}
  \end{teaserfigure}

\maketitle

\section{Introduction}

Visual icons are a popular tool for providing  quick, intuitive links to concepts. They are used in road signs, user interfaces, and in graphics. These icons serve to save space as they are smaller than text, provide universal access by avoiding language barriers, and reduce the cognitive workload for users through easy recognition and memorability. There is a long history of icon use on maps~\cite{MacEachren2004,Tyner2014}. 

Tactile maps are vital for people who are blind or have low vision to learn locations for independent travel and mobility~\cite{Banovic2013,Caddeo2006,Ungar1993role}. Maps are also widely used within education. However, perception and interpretation of these maps  is a complex task~\cite{Aldrich2002}.
Traditional tactile maps take the form of raised line drawings, using a combination of lines, textures and point symbols~\cite{Gill1973,Rener1993tactile}, often all raised to the same height. Due to the limitations of tactile perception~\cite{Heller2014psychology}, map features are indicated using braille letters or abstract symbols that need to be understood in conjunction with a legend, further complicating the already difficult task of tactile reading. 

3D printing is a new technology offering an alternative to raised line drawings. It provides more information through the third dimension and the tactual experience is more closely related to real-world experiences for people who are blind ~\cite{Heller2014psychology,Holloway2018accessible}. Accordingly, 3D printed maps have become a popular area for HCI research, e.g. \cite{Brule2016mapsense,Hofmann2022maptimizer,Kim2015,Lee2016finger,Shi2020molder,Taylor2016customizable}, however investigation of 3D tactile icons is in its infancy, with most research still focusing on abstract symbols~\cite{Gual2015improving,Simonnet2018maritime,Wabinski2022applying}. 

Preliminary work has suggested that some representational 3D tactile icons can be recognised by touch without use of a legend, reducing the cognitive effort required to interpret tactile maps and providing a format that is engaging for both blind and sighted users~\cite{Coughlan2022,Holloway20193D,Wang2022}. 
Such a solution would address BLV requests for standardised tactile map icons that can be used across a range of maps. 3D printing offers a practical method for iterative co-design and production of these icons.

This work addresses three fundamental challenges to the adoption of representational 3D icons for tactile maps:
\begin{enumerate}
    \item We do not know whether it is possible to remove the need for abstract point symbols on tactile maps by replacing them with a wide range of representational 3D tactile icons that can be understood more easily and minimise the need to refer to a legend.
\end{enumerate}
Next, a library of icons will need to be developed and verified for use on a wide range of maps. But:
\begin{enumerate}
  \setcounter{enumi}{1}
    \item We have only limited guidelines for the creation of 3D icons suitable for touch recognition by people who are blind or have low vision~\cite{Holloway20193D}. And;
    \item We need a practical methodology to co-design and test the efficacy of these 3D icons. 
\end{enumerate}

To address these challenges, we conducted two studies.  These considered icons for use on maps of parks and on maps of shopping precincts as representative and practically important case studies. In Study 1, an initial set of 186 icons were designed. The prototypes were tested with sighted people to quickly eliminate any designs that could not be adequately perceived by novice touch readers, and the top 86 icons were touch tested by at least ten sighted and ten blind people to confirm their suitability. The results revealed how the unique abilities of the two populations impacts on their understanding through touch. On this basis, a recommended methodology for touch testing was devised and used in Study 2 to further refine a set of recommended of 3D tactile icons or TactIcons. 

This work makes contributions in two distinct areas. 
\begin{enumerate}
\item Methodology:
\begin{itemize}
\item General principles for the  design of effective 3D tactile icons. 
\item A practical methodology for testing new 3D representational tactile icons that recognises the individual differences and strengths of both novice and congenitally blind touch readers.  
\end{itemize}
\item Icon Development:
\begin{itemize}
\item The first study to show that a wide range of 3D representational icons can be recognised by touch without prior exposure or reference to a legend, potentially making 3D maps easier to use for people who are blind or have low vision. 
\item A corpus of 3D printed icons for maps of parks, playgrounds and shops that have been pre-tested, shared at \url{https://www.thingiverse.com/thing:5841775}. Thirty three of these icons can instantly be understood by touch while a further 34 are tactually distinct and can be readily understood in conjunction with a legend. 
\end{itemize}
\end{enumerate}

It is hoped that this work will contribute to the creation of user-friendly tactile maps for people who are blind or have low vision, thereby supporting independent travel and educational success. The use of representational 3D tactile icons on maps has the added advantage of being useful and appealing to the general public, creating an inclusive experience for all. 
\section{Related Work}
\subsection{Visual Icons}
Icons are symbols to represent a concept or map feature with a physical resemblance to the object they represent (e.g. a tree), an associated object (e.g. a coffee cup for a cafe) or an associated emblem (e.g. a red cross for medical assistance)~\cite{MacEachren2004,Robinson1984elements}. 
A successful icon will depict an object that is easy to recognise, is strongly associated with its intended purpose or map feature,  can be easily discriminated from other icons and is memorable~\cite{Rogers1989icons,Tyner2014}.
This reduces the need to refer to a legend, which would otherwise add to  task complexity~\cite{Gobel2018,Netzel2017}.
Representational icons are generally more successful than abstract icons~\cite{Leung2002,Satcharoen2018}. 

Experienced designers develop new icons by first ideating on associations with the concept they are trying to convey, then searching existing image and icon repositories to get inspiration~\cite{Zhao2020iconate}. They also follow guidelines for icon design, e.g. \cite{Buhler2020universal,Google2021,ISO2010}, which are strongly based on visual perception and pictorial conventions. Recently, B\"uhler and colleagues proposed a more rigorous process for designing universal and intuitive pictograms~\cite{buhler2022UIPP} whereby consideration should be given to the full range of potential users. Testing with objective measures should be conducted with diverse members of the user population, with iterations based on their feedback~\cite{Abeysekera2001,buhler2022UIPP,Nolan1989}. Nevertheless, tactile access is not a consideration for the design of visual icons.

\begin{figure}
    \centering
    \includegraphics[width=0.95\linewidth]{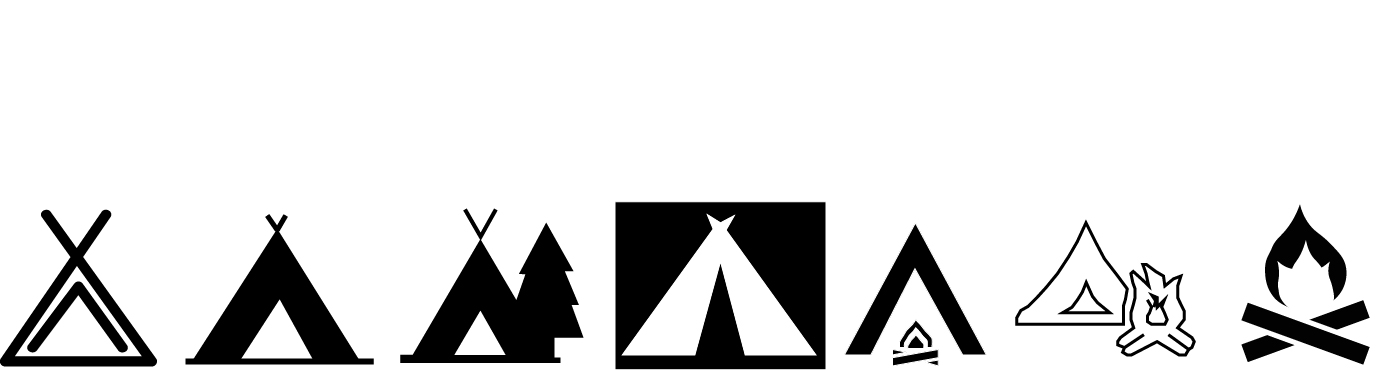}
    \caption{A variety of symbols used on print maps to represent camp grounds. There is little standardisation but all are effective for instant recognition.}
     \Description{Seven black-and-white icons using different line thicknesses, shading and shapes to depict a tent, a tent with a tree, a tent with a campfire or a campfire.}
    \label{fig:camping}
\end{figure}

As illustrated in Figure~\ref{fig:camping}, the need to ensure that icons can be easily understood does not constrain individual or stylistic choices, with a  wide variety of icons possible for the same map feature, all of them effective. 

\subsection{Symbols on Tactile Maps}
Haptic perception has a much lower resolution compared to visual perception~\cite{Hatwell2003touch,Heller2014psychology} and the graphic's elements must be explored sequentially rather than perceived in parallel, increasing cognitive load~\cite{Millar2003how}. 
Given these differences between print and tactile graphics, the content of tactile maps is usually simplified, with lines, symbols and spacing re-designed for easy tactual discrimination.  

\begin{figure}[htb]
    \centering
    \includegraphics[width=0.5\textwidth]{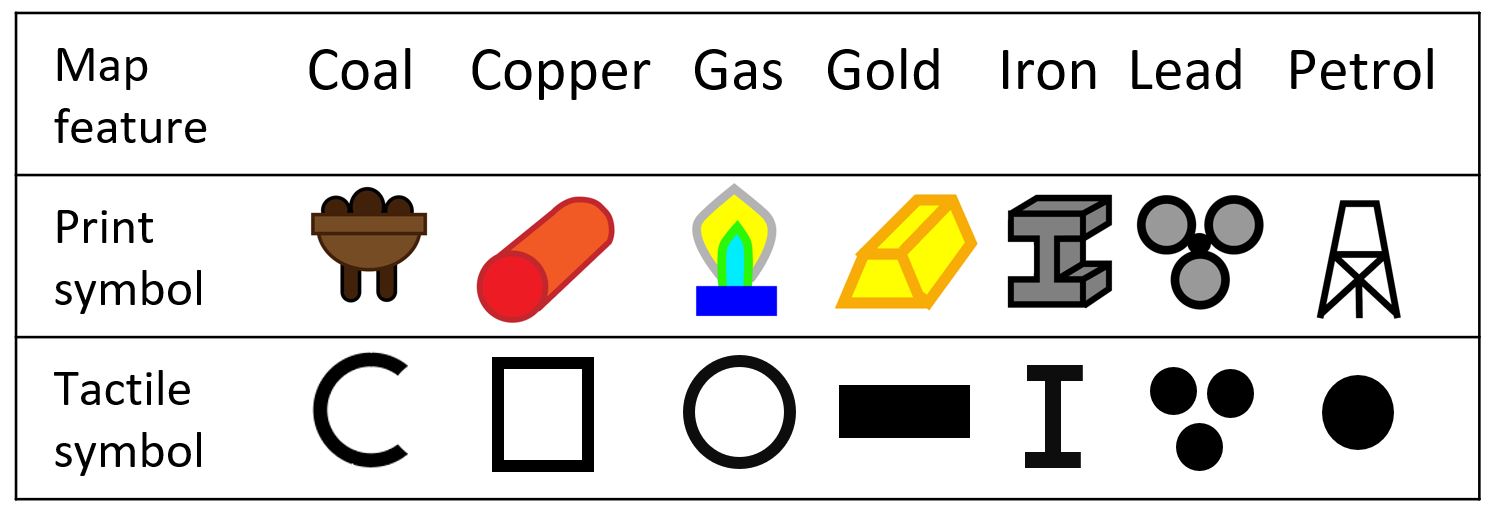}
    \caption{Representational print icons and corresponding abstract tactile icons for land use, adapted from an example map in the  Guidelines and Standards for Tactile Graphics by the Braille Authority of North America~\cite{BANA2010guidelines}.}
    \Description{Print and tactile symbols for coal, copper, gas, gold, iron, lead and petrol. The print symbols use colour and detail for representational icons such as a copper pipe, gas flame and gold bar. The corresponding tactile symbols are simplified shapes, for example a square for copper, circle for gas and rectangle for gold.}
    \label{fig:BANA}
\end{figure}

Commonly, tactile map features are labelled using a two- or three-cell braille symbol that is then explained in an accompanying legend~\cite{BANA2010guidelines,Eriksson2003tactile,RoundTable2022guidelines}. 
Another approach is to use highly simplified symbols, with a limit of no more than 10 to 15 distinct symbols on a single map~\cite{Lobben2012,Rowell2003world}. On rare occasions, attempts are made to use tactile symbols related to the original print symbols (Figure~\ref{fig:BANA}), however problems have been found with recognition of icons as simple as a cross or a star~\cite{CBA2003report}. Regardless, the use of both braille symbols and point symbols relies on reference to a legend, adding to the cognitive complexity of reading and understanding tactile maps. 

Blind and low vision touch readers commonly ask for a standardised symbol set for tactile mapping~\cite{Holloway20193D,Rowell2003feeling,Rowell2003Taxonomy} so they won't need to decode the symbols for every new map they encounter, thereby reducing cognitive load. Repeated attempts have been made to standardise tactile map symbols, e.g. ~\cite{ADON1986,James1975handbook,Lambert1989,Lobben2012}, however none of these systems have been adopted by the wider tactile graphics production community. 
Instead, it has been argued that standardisation of symbols for raised line maps is not achievable~\cite{Eriksson1999,Tatham2001} due to several factors.
Hundreds of distinct symbols are required for even a single type of map~\cite{Bandrova2001,MacEachren2004}, but discrimination between even simple tactile shapes is difficult~\cite{Austin1952,CBA2003report,Gill1973} and only a limited number of tactile symbols can be distinguished from one another on a single map~\cite{Cole2021,Lobben2012,Rowell2003world}.
Furthermore, designers of tactile graphics use a variety of different CAD software and do not have a shared repository of graphic components~\cite{Rowell2003Taxonomy}. As we shall discuss, the introduction of 3D tactile icons may make standardisation may be more feasible if a wide range of representational icons are easier to understand.

\subsection{3D Printed Tactile Maps}
Compared with tactile graphic representations, which require interpretation using visual conventions such as outlines and perspective, 3D objects are easier to recognise by touch due to cues such as volume, size, texture and weight~\cite{Klatzky1985identifying,Klatzky1993haptic,Lederman1990visual}. There is early evidence that 3D printed models, which provide information about volumetric shapes, are likewise easier to understand than tactile graphics~\cite{Holloway2018accessible}. 

Simonnet et al. found that a small number of abstracted 3D shapes for maps could be recognised at a scale of 5mm~\cite{Simonnet2018maritime}. Gual, Puyuelo and Lloveras  found that 3D abstract symbols are more memorable than 2.5D tactile symbols~\cite{Gual2014three,Gual2015improving,Gual2015effect}. These studies were consistent with the general practice of using abstract shapes and textures on tactile graphics so that they can be readily discriminated by touch, e.g.~\cite{ADON1986}. However, 3D printing allows for the introduction of tactile symbols that are more iconic, reducing the need for the reader to consult a legend.

Following a comparative study showing that representational map features such as stairs and seats made 3D printed maps easier to understand than 2.5D tactile maps~\cite{Holloway2018accessible}, we evaluated a tactile map with nine representational 3D icons and provided preliminary guidelines for their design~\cite{Holloway20193D}. Some, but not all, of the icons could be recognised without reference to a legend. Wang et al. went on to create eight representational 3D icons based on these guidelines and others~\cite{Gual2014three,Gual2015effect,Holloway20193D} but did not evaluate their usefulness beyond the fact that they were understandable when used on a 3D map~\cite{Wang2022}. Systematic testing of such icons is required, along with more detailed design guidelines.  

Interactive audio labelling offers another option for accessible mapping, and has received much recent attention~\cite{CavazosQuero2019,Doi2010,Ducasse2016,Gotzelmann2016,Palivcova2020,Taylor2016customizable,Wang2009}. Evaluation is generally positive, however audio labels cannot be used by people who are deafblind. Thus, audio labelling should be used in conjunction with (rather than as a replacement for) tactile solutions.

Here, we demonstrate that the adoption of 3D printing for tactile accessibility introduces the possibility of a wider set of tactile symbols that can be more readily recognised and remembered than is possible with 2.5D tactile symbols. While it may still be impossible to impose a definitive set of tactile icons for use in all tactile maps, we can offer a set of user-tested icons that are easily distributed as freely downloadable 3D printing files (serving as a database as conceptualised by Rowell and Ungar~\cite{Rowell2003Taxonomy}) as well as offering guidance on how to design additional 3D tactile icons. 
\section{Methodology}
\begin{figure*}[htb]
    \centering
    \includegraphics[width=0.8\textwidth]{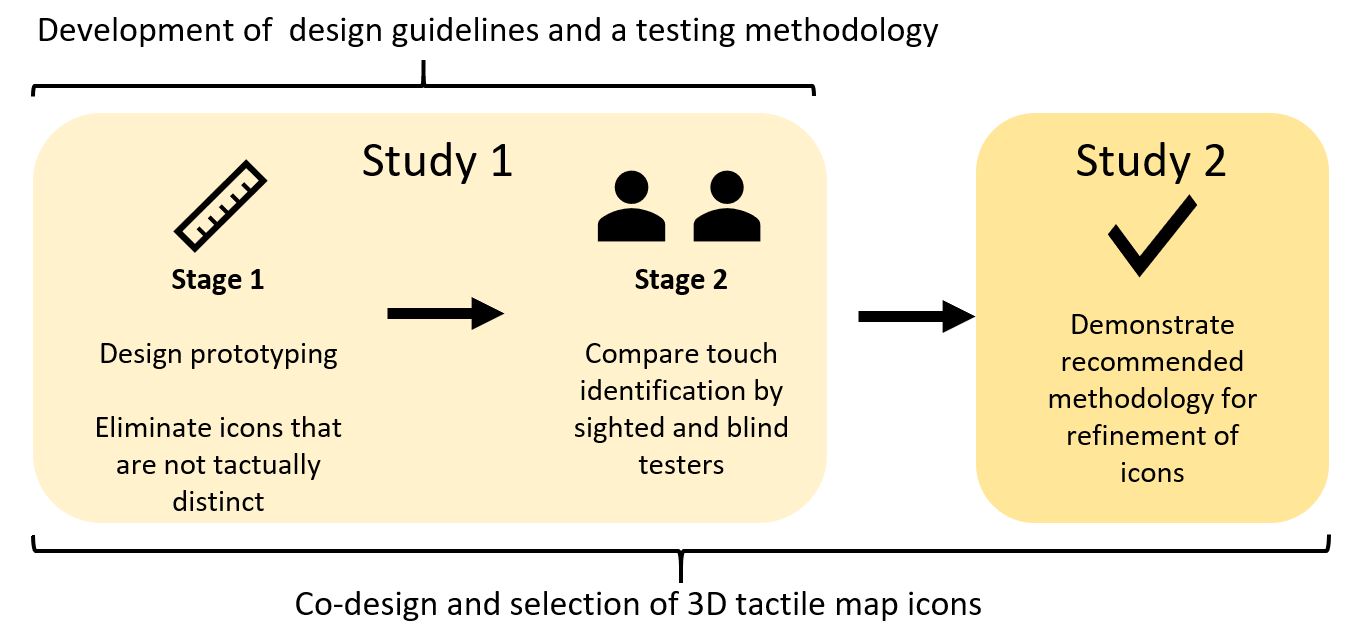}
    \caption{Aims and outputs across this paper's two studies.}
    \Description{Flow chart with two boxes for Study 1 and Study 2.
Study 1: Stage 1. Design prototyping & eliminate icons that are not tactually distinct. Arrow to Stage 2. Compare touch identification by sighted and blind testers
Study 2: Demonstrate recommended methodology for refinement of icons. 
Study 1 achieves development of design guidelines and a testing methodology. 
Studies 1 and 2 involve Co-design and selection of 3D tactile map icons.}
    \label{fig:studies}
\end{figure*}

This work was conducted across two studies (Figure~\ref{fig:studies}). 
Study 1 involved two stages. In stage 1, a large initial set of prototypes was co-designed with a congenitally blind touch reader then touch tested by sighted participants to quickly eliminate any designs that were perceptually indistinct. 
In stage 2, the remaining icons were tested with an additional five sighted people and ten blind people. This additional testing gave a larger and more appropriate testing pool and allowed a comparison of the two groups' abilities  to understand 3D icons. 
This comparison informed the formulation of a recommended methodology for touch testing 3D icons, addressing  Research Challenge 3. This methodology was then applied in Study 2 evaluating a selection of 3D icons that had been revised based on feedback from the touch testers in Study 1. Both studies thereby served to develop and test 3D icons for use on tactile maps and provided insights for the creation of design guidelines for further 3D icons, addressing Research Challenge 2.  

Sighted people were chosen for stage 1 testing because, as novice touch readers, they require icons to be tactually distinct in order to recognise them. Icons that can be recognised by sighted people therefore  allow for successful recognition not just by skilled touch readers but also people within the blind and low vision community who have recently acquired blindness, those who mainly rely on large print or audio formats, or people with age-related blindness and declining tactual acuity.  Sighted people were also chosen as a convenient sample who can easily be accessed for rapid prototyping. 

Co-design is an important tool for inclusive research practice ~\cite{Holloway2021disability,Strnadova2022}. At all stages of the research, participants were encouraged to make suggestions for improvements or to devise new icons for the same map features. These comments were recorded and coded for analysis and the most promising or interesting suggestions were also used to make minor revisions or introduce new icons into the testing pool. 

\section{Study 1: 3D Icon Design and Testing}
Study 1 aimed to create, refine and evaluate 3D tactile map icons using an iterative co-design process. A secondary aim was to compare the abilities of blind and sighted touch testers in their recognition of 3D tactile icons by touch, in order to determine if and when the use of sighted touch testers is appropriate in the development of tactile materials.

Work began with the iterative design and testing of a large corpus of tactile icons. After pre-testing by 3-5 sighted participants in stage 1, the icons were either rejected or continued to stage 2 for further touch testing by a total of 10 sighted and 10 BLV people. Throughout both stages, icons were refined and new icons were added for stage 1 testing, based on suggestions from the touch testers. 

\subsection{Materials}
\subsubsection{Subject Matter}

Tactile icons may need to be designed for a broad variety of maps and locations such as indoor maps, local neighbourhood maps, road maps, and thematic maps. We chose to focus on two types of maps relating to concurrent projects requested by members of the BLV community: parks with playgrounds and shopping precincts.  

Park and playground maps were requested by an O\&M professional and vision specialist teachers working with people who are BLV. An initial list of park and playground icons was constructed by examining the first 20 unique maps to appear in a Google image search for ``playground map''. 
Map features were considered for inclusion if they appeared in 25\% or more of the maps. 

An initial list of 20 shops was requested by a blind child learning O\&M skills in her local neighbourhood.  
A further 18 shops thought to be common or important were added to this list for prototyping of icons.  

Some of the chosen map features -- a library, coffee shop, entrance and toilets -- align well with the top 20 points of interest identified for inclusion on accessible indoor and university maps by Papadopoulos and colleagues~\cite{Papa2016}. They also suggested stairs, which was added to the list of park icons. Their other icons were not relevant to this study as they were specific to indoor environments and a university context.

\subsubsection{Scale}

Icons were restricted in size to 20mm\textsuperscript{3}, as this is similar in size to a 2 or 3 braille cell abbreviation, which occupies approximately 22mm$\times$11mm including surrounding space~\cite{PharmaBraille} and is recommended for tactile graphic maps~\cite{BANA2010guidelines,Eriksson2003tactile,RoundTable2022guidelines}. This small scale will assist in keeping maps at a usable size, being mindful that touch reading requires everything to be within arm's reach.

All icons were placed on a 20mm diameter disc, which would assist with identifying which way up the icon should stand, as well as preventing the fingers from exploring the icon from underneath, similarly to when the icon is fixed to a map. A minimum thickness of 2mm was used for most components within the models to ensure adequate strength and durability.

\subsubsection{Design}
Following a similar process to that used by designers of print icons~\cite{Zhao2020iconate}, initial design ideas were generated through free associations and reference to existing icons: in print; prior research~\cite{Holloway2018accessible,Holloway20193D}; and the tactile symbols for communication developed by the Texas School for the Blind and Visually Impaired~\cite{TSBVI}. All practical ideas for icons were included without pre-judgement as to whether the concepts they used would be familiar to people who are blind. 

\begin{figure*}[htb]
    \centering
    \includegraphics[width=13.5cm]{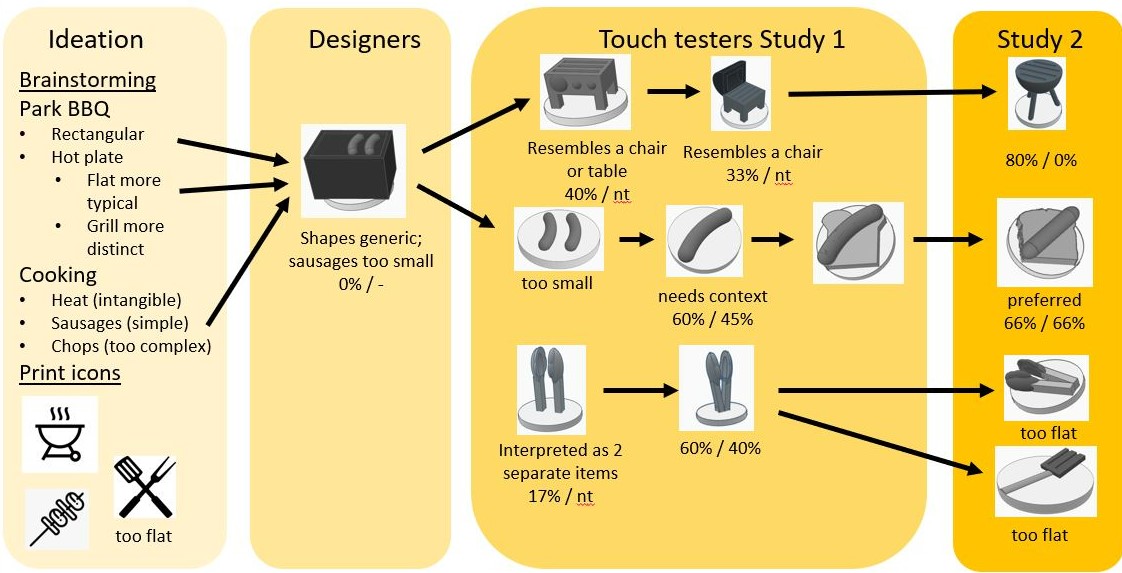}
    \caption{Icons developed to represent a barbeque area based on brainstorming and reference to print icons. Minor adjustments such as adding and refining the scallops on the tongs are not shown here. Percentages indicate recognition accuracy for sighted/BLV touch testers, where `nt' = not tested.}
    \Description{Flow chart with four boxes from left to right. 
Box 1 ideation. Brainstorming: Park BBQ--rectangular; hot plate (flat more typical, grill more distinct). Cooking--Heat (intangible); Sausages (simple); chops (too complex). Print Icons: round BBQ with heat waves, shishkabob, and crossed spatula and BBQ fork. 
Box 2 designers. 3D icon of rectangular BBQ with two sausages on a flat hotplate. Arrows from “rectangular”, “flat” and “sausages” in the ideation box. Arrows to BBQ and sausage icons in box 3.  
Box 3 touch testers study 1. 3D icon of rectangular BBQ with grill top and knobs at front (resembles a chair or table. 40\%/nt. Arrow to icon of square BBQ with raised lid (resembles a chair, 33\%/nt). Arrow to round BBQ in box 4. 
Next line 3D icon of two bent sausages (too small). Arrow to icon of one bent sausage (needs context, 60\%/45\%). Arrow to icon of bent sausage on stylized slice of bread. 
Next line 3D icon of straight upright tongs (interpreted as 2 separate items. 17\%/nt). Arrow to slightly parted upright tongs (60\%/40\%). Arrows to flat tongs and spatula in box 4. 
Box 4 study 2. Round BBZ (80\%/0\%). Next line straight sausage on naturalistic slice of bead (preferred, 66\%/66\%). Next line slightly parted tongs lying down (too flat). Next line spatula lying down (too flat). 
}
    \label{fig:BBQ}
\end{figure*}

The initial prototype icons were created by a sighted accessible formats designer in consultation with a congenitally blind collaborator. Consideration was given to prior guidelines for the design of tactile graphics and icons, particularly the need to simplify~\cite{BANA2010guidelines,Barth1988,Challis2000,Edman1992,McLennan1998} and to position the most important features at the top of the icon for easy touch access~\cite{Holloway20193D}. Once touch testing commenced, all practical suggestions for refinements or new icons were produced and introduced into the testing pool, as illustrated in Figure~\ref{fig:BBQ}.

The icons were designed for production on a FDM 3D printer, as these are most widely available. Extreme overhangs were avoided so that all icons could be printed without supports or assembly of separate parts, with the exception of icons for a swing and tap.  
All printed icons were checked for sharp points and some minor filing/sanding was conducted, mainly underneath overhangs or to remove printing anomalies, for touch reading safety and comfort. 

A total of 186 icons were developed for 84 map features across the two map types. 

\subsection{Participants}

A total of 46 people contributed to Study 1. Among the thirty five sighted participants, the 16 men and 19 women were aged from 13 to 70 ($\bar{x}=37$, sd=15). One was in the early stages of vision loss and uses large print, while the remainder had normal or corrected vision and were able to read standard print. One potential participant was excluded from the study due to self-reported de-sensitivity of the finger pads. 

Eleven BLV people contributed to Study 1--two men, eight women and one non-binary individual. All are braille-literate touch readers and they were aged from 29 to 67 years old ($\bar{x}=48$, sd=16). 

The participant profiles across Studies 1 and 2 are given in Table~\ref{tab:participants}.

\subsection{Procedure}
\subsubsection{Touch Testing}
Some touch testing sessions were conducted at public events, when results were collated to speed up the testing process. All other sessions were held individually and in person with the exception of one session with a blind participant that was completed remotely over Zoom, with the materials sent out in advance. This method was not practical for more participants due to the large number of icons that needed to be 3D printed. 

When touch testing with sighted participants, the tactile icons were presented out of sight using their preferred method -- in a box with a curtain over the open front, or held in the hands without looking. 

Participants were told whether each icon was designed for a map of a shopping district, park or playground, as this would ordinarily be known when the icons are used on a map with a title, and because categorical information has been found to assist in tactile identification~\cite{Heller1996tactual,Picard2014}.
Participants were asked to name the object and then, if they were correct, to suggest what it might represent on a map. Correct answers were accepted even if it was not the first guess. After looking at the icon, they were also invited to make suggestions for improvements or alternative symbols to represent the same map feature. 

Due to the large number of icons in stage 1 and iterative design method, most sighted participants tested only a subset of the 3D icons. If they had already tested some icons in a prior session, these were removed before random selection of further icons began. As documented in Table~\ref{tab:participants}, sighted participants tested from 1 to 182 icons ($\bar{x}=60$, sd=56) per person. All blind participants tested a full set of icons selected for stage 2 and some additional iterations, ranging from 51 to 89 icons tested per person ($\bar{x}=56$, sd=11).

After testing all of the icons, blind participants were re-tested on the icons they had been unable to recognise in order to determine whether the icons were both tactually distinct and memorable, so that they could be recognised in conjunction with a legend.

\subsubsection{Icon refinement, rejection and selection}
Icon testing, refinement and selections was conducted on a rolling basis, with a constantly changing pool of icons to be tested.

Suggested modifications to icons were classified as minor or major according to re-testing response. 
For minor changes, testing continued with the initial responses included. For example, adding small scallops to a set of barbeque tongs was said to be ``a tiny bit better'' and therefore classified as a minor change. 
For major changes or new icons, the new icon was added to the testing pool and the original icon was either discarded (if the modification improved on an identified problem) or otherwise retained for continued testing. 
For example (Figure~\ref{fig:BBQ}, changing the barbeque tongs from straight to angled ``makes much more sense'' and was therefore classified as a major change, with testing of the straight tongs discontinued. Changing the orientation of the tongs to flat was classified as a major change, with continued testing of the original upright tongs along with testin of the new flat tongs.   

All icons were first pre-tested with sighted participants (stage 1). If an icon object had a 0\% success rate after the  first three participants, it was excluded from further testing. All other icons were tested by a minimum of five sighted participants, when further selection was conducted. Icons with an identification rate of 20\% or less were rejected. In addition, where several icons represented the same object, the icon with the highest accuracy rate was chosen for further testing in stage 2. However, where the same concept was represented using different objects, both potential icons were included for further testing if they had an accuracy of 40\% or above. 

After initial testing of all icons by five sighted participants in stage 1, eight icons with identification rates below 50\% were scaled up by 200\%, to determine the impact of scale on recognition of the icons. The enlarged icons were tested with a minimum of five people (sighted or blind) who had not been exposed to the smaller icon. 

\subsection{Results}

\subsubsection{Stage 1 Icons Accuracy}
From the initial pool of 120 shop icons and 63 icons for parks and playgrounds, 29 were discarded early in the testing process after failing to be correctly identified by the first three participants to test them. 
A further 12 icons were discarded because the identification rate was 20\% or less after being tested by at least five sighted people.   
Finally, 50 icons were discarded because they were based on the same concept as another icon with a higher recognition rate, or because recognition was below 40\% and there was another icon for the same map feature with better recognition. For example, a sunflower in a vase (60\% accuracy) was included for testing and a rose in a vase (50\% accuracy) was omitted. 
This selection process resulted in 86 icons to represent 74 map features. There were ten map features for which no icons had been successful, such as a hairdresser and antique store. 

Many of the failed designs had small details that could not easily be detected by touch, such as a  sewing machine and scissors. Icons that were laying down and therefore essentially flat were most prone to this problem. Likewise, small holes were often not detected, such as a face port on a massage table. 
Other failed designs were so simple that their shape was not distinguishable, as was the case with a pill bottle and cotton reel. 

\subsubsection{Stage 2 Icons Accuracy}
\begin{figure*}[htb]
    \centering
    \includegraphics[width=0.8\textwidth]{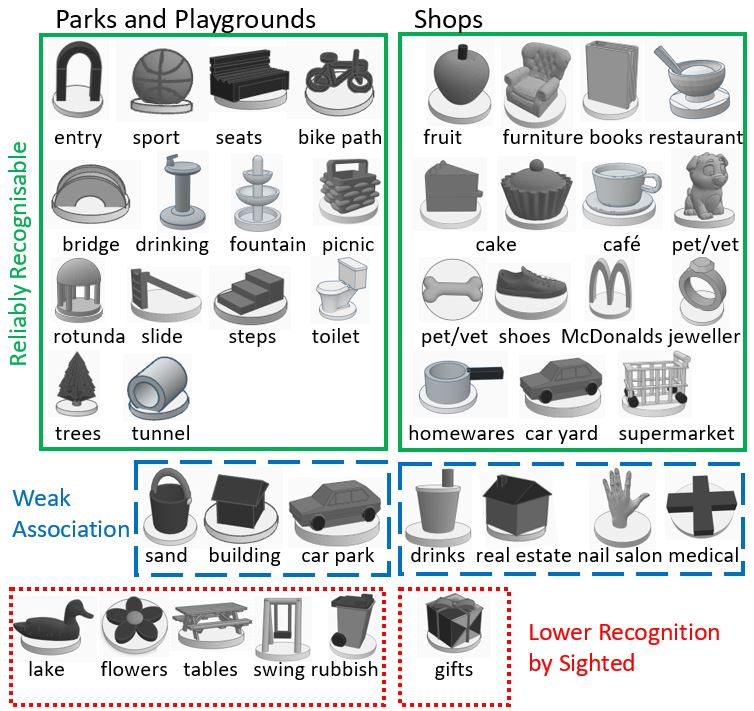}
    \caption{Icons for park and playground maps (left) and shop maps (right) recognised by 80\% or more of blind touch testers in Study 2. Icons with a dashed blue outline had an association rate less than 80\% by blind testers. Icons with a dotted red outline had a recognition rate less than 80\% by sighted testers.}
    \Description{Reliably recognisable icons for parks and playgrounds: archway (entry), ball (sports), bench seat (seats), bicycle (bike path), arched bridge (bridge), drinking fountain, 3-tiered fountain, picnic basket (picnic), rotunda, slide, 3 steps (steps), toilet with cistern (toilet), pine tree (tree), tunnel. 
Reliably recognisable icons for shops: apple (fruit), armchair (furniture), book, bowl with chopsticks (restaurant), slice of cake with cream filling (cake), cupcake with patty pan and cherry on top (cake), cup and saucer (café), cartoon dog with short snout and upright tail (vet/pet), dog bone (vet/pet), loafer (shoes), McDonalds M (McDonalds), ring with jewel (jeweller), saucepan (homewares), sedan car (car yard), supermarket trolley (supermarket). 
Weak association icons for parks and playgrounds: bucket with handle up (sand), building with roof (building), sedan car (car park). 
Weak association icons for shops: disposable cup with lid and straw (drinks), building with roof and chimney (real estate), outstretched hand (nail salon), flat cross (medical). 
Icons with lower recognition by sighted for parks and playgrounds: duck (lake), daisy head (flowers), picnic table with seats (tables), swing in a-frame (swing), wheelie bin (rubbish). Icon with lower recognition by sighted for shops: package tied with string and bow on top (gifts).
}
    \label{fig:top40}
\end{figure*}

Further touch testing in stage 2 with blind and additional sighted participants confirmed that many of the pre-selected tactile icons could indeed be recognised by touch without reference to a legend.
Forty two icons were correctly recognised by 80\% or more of blind participants.  
As seen in Figure~\ref{fig:top40}, the majority of these icons are based on simple but distinct concepts, such as a cupcake with a distinctive patty pan. Full results for the successful icons are given in Tables~\ref{tab:recognisable} and \ref{tab:learnable}.

Twenty nine of these icons could consistently be identified and correctly associated by at least 80\% of both sighted and blind touch testers. Six of the remaining successful icons had recognition rates of only 50-70\% by sighted touch testers.  
These icons were somewhat more complex or with less well-defined features that novice touch readers may have had difficulty detecting.  

In addition, the participants struggled to associate seven of the well-recognised icons with a map feature. For example, an outstretched hand was used to refer to a nail salon. While the hand was instantly recognisable, in the context of shops it could also be interpreted as signifying help or a glove shop.  
A legend would be needed to ensure correct association for these icons. The association for some of the successful icons also depends on context and they could potentially be applied to other types of maps beyond parks and shopping precincts. After an icon was unable to be identified on first exposure, recall by BLV touch testers was very high on the second attempt, ranging from 75\% to 100\% for individuals ($\bar{x}=90\%$, sd=9), and at an overall a rate of 87\%. This suggests that 3D tactile icons are memorable, reducing the need to refer to a legend more than once.

\subsubsection{Scale}
\begin{figure*}[htb]
    \centering
    \includegraphics[width=12cm]{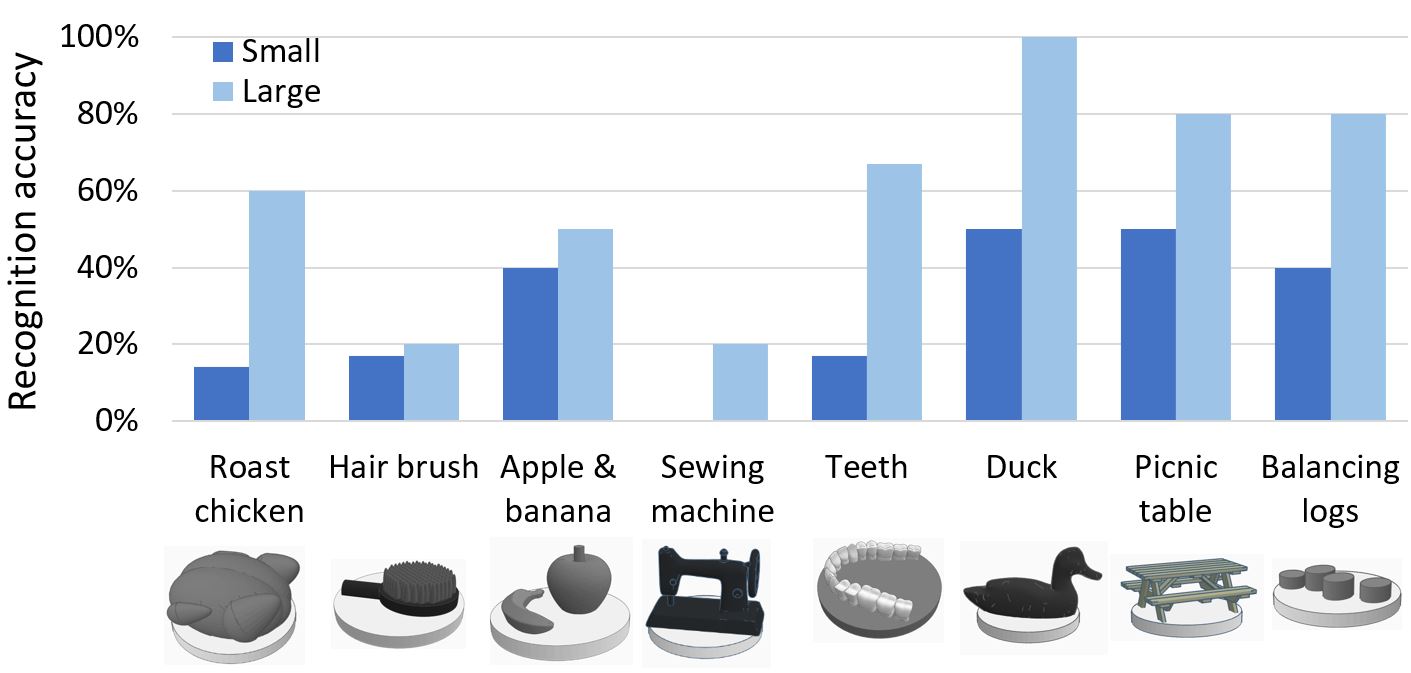}
    \caption{Touch recognition accuracy for small ($20mm^3$) and large ($40mm^3$) icons by sighted people.}
    \Description{Bar graph charting recognition accuracy for small - large icons. Roast chicken: 14\% - 60\%. Hair brush: 17\% - 20\%. Apple and banana: 40\% - 50\%. Sewing machine: 0\% - 20\%. Teeth: 17\% - 67\%. Duck: 50\% - 100\%. Picnic table: 50\% - 80\%. Balancing logs: 40\% - 80\%.}
    \label{fig:scale}
\end{figure*}

As shown in Figure~\ref{fig:scale}, recognition of enlarged icons ($40mm^3$) was consistently better than recognition of the same icon at the original small size ($20mm^3$). The increase in recognition accuracy was greatest for icons that are a distinctive shape to allow recognition but too complex for tactile recognition by novice touch readers at a small scale. 
Recognition remained low for the hair brush, with participants reporting that the texture of the bristles was wrong, and for the sewing machine, which is a complex and relatively uncommon object. 
Overall though, these findings confirm that tactile recognition of most 3D icons can be made easier by increasing their size (where the map size permits), in keeping with previous studies of 3D abstract icons~\cite{Chen2010}.

\subsubsection{Group and Individual Differences in Recognition and Association}

Overall, accuracy was slightly higher for parks and playgrounds (56\%) than shops (52\%), with some participants reporting that they found park icons easier because the context was more specific, whereas a shop could sell anything. 

Accuracy for some icons was much higher for sighted than blind participants. There was a difference in recognition accuracy of at least 40\% for 11 icons. 
Discussion revealed two reasons for the low recognition rate by blind participants: they were either less familiar with these objects; or the icon had used archetypes that are common in visual communication but are not accurate representations of the real object. The cactus in a pot illustrates both these points: blind people are much less likely to touch a spiky cactus than other types of plants; and a straight cactus with folds is a common print representation but is atypical of most real-life cacti. 

Conversely, a further eight icons were recognised more accurately by blind than sighted participants, with a recognition rate at least 30\% higher. Most of these icons, such as a swing in an A-frame and a power plug, have small details that require fine tactile exploration and perception. Tactile discrimination was observed to depend on both tactile acuity (ability to perceive the object) and use of effective tactile exploration strategies such as using two hands, moving the fingers pads over all of the object, and deliberate actions such as poking, pinching and squeezing.  

Individual accuracy rates varied greatly, with participants who tested at least five icons ranging in accuracy from 21\% to 90\% ($\bar{x}=62.7, sd=13.6$).
One blind participant had a very low overall accuracy rate of only 21\% overall, despite being able to describe the model parts and detect small details. For example, they described a round container that you can put things in, with a long stick coming out horizontally from the side at the top, but they could not identify it as a saucepan. Similarly, a sighted participant was able to describe what they were feeling so that a bystander could correctly guess the object, but they could not guess for themselves. This highlights the importance of individual differences in mental imagery ability. 

Clearly, some of the icons shortlisted in stage 1 were inappropriate for use by blind touch readers due to the reliance on concepts that they may not be familiar with. 
Input from congenitally blind touch readers is required to ensure that 3D icons are suited to their target audience. In addition, sighted people can help to identify any 3D icons that are not tactually distinct enough for novice touch readers. 

\section{Implications}
The results from Study 1 provide helpful observations about the characteristics of 3D icons, their impact on ease of understanding, and how such icons can be designed and evaluated. 
\subsection{Object Identification}
Correct recognition of a 3D icon by touch was observed to rely on three factors:
\begin{enumerate}
    \item \textbf{Tactile perception:} Skilled blind touch readers may have superior tactile perception due to remapping of the brain~\cite{Sathian2000} and use of effective tactile reading strategies to support improved tactile picture recognition~\cite{DAngiulli1998blind,DAngiulli2000guided,Lederman1987hand}. We observed that touch readers were much more likely to use multiple fingers on both hands at once to explore the icon from all angles, whereas sighted people were more likely to use just one finger. Skilled touch readers also reported using deliberate manual exploration strategies, such as gently squeezing an icon to feel for holes or gaps.
    \item \textbf{Mental imagery:} The ability to visualise an object based on a description or touch access to its parts can differ markedly from one individual to the next~\cite{Galton1883,Kosslyn1984individual,Zeman2020} and has previously been found to impact recognition of raised line drawings~\cite{Lebaz2011haptic}. A small number of our touch testers, both sighted and blind, exhibited difficulty with mental imagery and would be unlikely to benefit from tactile icons regardless of how easily they can be perceived. 
    \item \textbf{Familiarity with the object:} Blind people may be unfamiliar with some objects, especially those that are extremely large or small and cannot be touched. This was observed as an important factor in tactile icon recognition, for example with icons such as the skateboard half pipe and cactus, and aligns with a previous study in which congenitally blind participants had more trouble recognising print icons than those who had visual memory~\cite{Chen2010}. Age and cultural factors can also impact familiarity with objects and their associations. For example, older participants more easily recognised round metal trash cans, whereas younger participants preferred modern rectangular wheelie bins. 
\end{enumerate}

Object identification requires not just recognition of the object, but also correct association of the recognised object with the intended map feature.
Study 1 did not reveal significant  differences between blind and sighted participants in association once the object was recognised.

\subsection{Proposed Methodology for Touch Testing 3D Icons}
\begin{figure*}[htb]
    \centering
    \includegraphics[width=12cm]{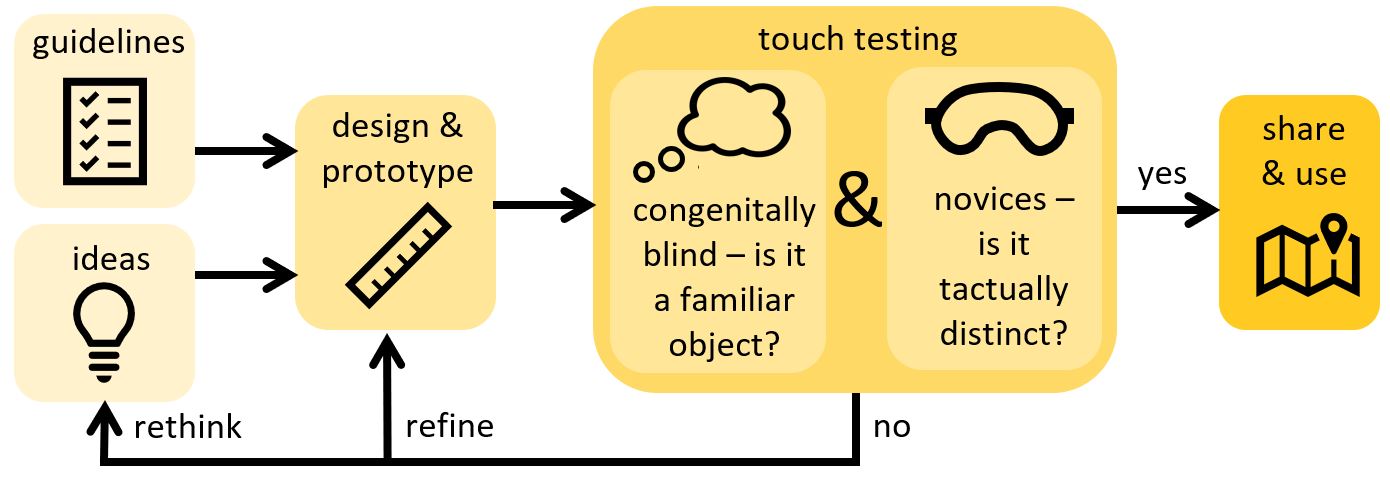}
    \caption{Proposed method for the design and testing of tactile icons. Design is based on ideas (from print icons or associations) and adheres to design guidelines. People who are congenitally blind check whether the icons are based on concepts that are familiar to them and novice touch readers check whether icons are tactually distinct. Feedback from both groups determines whether icons are accepted for use or should be refined or replaced.}
    \Description{Flow chart beginning with (1) guidelines and (2) ideas. Arrows to (3) design & prototype. Arrow to (4) touch testing by (a) congenitally blind touch testing - is it a familiar object? AND (b) novice touch testing - is it tactually distinct? If no, arrows labelled "rethink" back to (2) ideas and labelled "refine" back to (3) design & prototype. If yes, arrow forward to (5) share and use.}
    \label{fig:testing_method}
\end{figure*}

Given the factors that impact understanding of 3D tactile icons by touch, a methodology for testing 3D icons is proposed: \textbf{icons should be touch tested by both novice touch readers for tactile distinctiveness, and also with congenitally blind people to ensure that they are familiar with the concepts on which the icon is based} (Figure~\ref{fig:testing_method}). The novice touch readers may either be sighted adults (without using their vision to identify the icons), or people who are blind or have low vision and do not have well-developed touch reading skills. Children should not be used as novice touch readers, as tactile perception is influenced by age, with greater tactile acuity in children~\cite{Gescheider1994} and with smaller fingertip size~\cite{Peters2009}.  The icons should also be touch tested by people who are congenitally blind or who do not have a strong visual memory. Testing should not be conducted with people who have poor mental imagery abilities. Pre-screening may be conducted using a test of mental imagery generation~\cite{DiNuovo2014,Pearson2013}, however many of these tests are  inappropriate for people who are BLV. Difficulty with mental imagery should also become evident during the process of touch testing icons, characterised by an identification accuracy rate outside the 95th percentile in spite of an ability to accurately describe the physical properties of the icon.  

\textbf{This proposed method for testing the efficacy of 3D tactile icons addresses Research Challenge 3.}

\section{Study 2: Refinement}
Study 2 implemented and verified our proposed methodology for tactile icon testing in order to refine icons based on suggestions from the blind touch readers in Study 1. Novice adult touch readers checked for tactile distinctiveness, and association was verified by congenitally blind touch readers. 
\subsection{Materials} 
Twenty three icons were chosen for refinement based on suggestions made by participants in Study 1. A further 17 suggestions for new icons were also created.  
The corresponding icons tested in Study 1 were also included for comparison, making a total of 70 icons for 24 map features to be tested in Study 2. 
\subsection{Participants}
Testing was conducted with three novice touch readers and three blind people with limited visual experience. 
The novice touch readers included one person with full vision, one person with deteriorating vision who uses large print and has not yet begun exploring tactile options, and one person who is legally blind but mainly uses large print at close range. They were aged from 49 to 63 years ($\bar{x}=57$, sd=7).
The congenitally blind touch readers all had minimal visual experience and were lifelong braille users. They were aged from 29 to 61 years ($\bar{x}=49$, sd=17).

Both groups included one man and two women. Both groups also included two participants from Study 1 and one person who had not been exposed to any of the earlier test materials.   
\subsection{Procedure}
All but one of the participants were presented with icons one at a time and asked to use only their sense of touch to identify the object and what it would represent on a map. All shop icons were presented together, as were all park and playground icons. If two icons were variations of the same object, the new design was presented first and the old design was presented immediately after for a direct comparison. Icons representing the same map feature but using a different object were not presented consecutively. The order of presentation was otherwise randomised. Once all icons for a particular map feature had been explored, participants were asked which they preferred. 

One congenitally blind participant suffered from mild memory loss and experienced difficulty when following the standard procedure for testing of park and playground icons. They were unlikely to be able to complete testing this way for the shop icons, which are generally harder to identify. 
The testing procedure was therefore adjusted -- all icons for a single shop were provided at once along with the name of the shop. This additional contextual information allowed the participant to more easily identify the icons so they could decide which was their favourite. This adjusted procedure achieved the same goal of identifying which icons are most readily recognisable by a congenitally blind touch reader, whilst being respectful of individual needs and ensuring that they could contribute to the research.

\subsection{Results}
\subsubsection{Co-design}
Comments regarding difficulty identifying the icons, along with suggestions for refinements or new icons, were recorded throughout Studies 1 and 2 then coded. In total, 409 comments were captured, with 74\% of recorded comments from blind participants. The most important factors in identification of icons were tactile perception (30 comments),  familiarity of objects and concepts (28 comments), distinctive or unique shapes (16 comments) and perceived realism (13 comments). Notably, all of the comments regarding familiarity and concepts were made by blind participants, with the exception of one comment by a sighted teenager. 
Overall, there was a tension between the desire for realism or additional details (51 comments) and the need to simplify (12 comments) or exaggerate features (11 comments) so that the icons would be tactually distinct (49 comments). 
All of the suggestions to add more details (8 comments) or place objects flat on their side (11 comments), which would add to difficulty with tactual perception, came from blind participants who are generally more skilled touch readers. 

\subsubsection{Selection of updated icons}
Results from Study 2 were helpful for evaluating the new and revised icons. Four new icons had recognition rates over 80\% for both novice touch readers and congenitally blind testers, and were preferred over the existing icons. 
A fifth icon gained 100\% recognition but there was no preference over the existing icon for the same map feature. These five icons can be recommended for use as they are reliably recognisable. 

A further four revised icons proved to be slight improvements over the prior icons depicting the same object. For the remaining map features, either the original icon was preferred (4 icons), there was no clear winner (8 icons) or further refinements were needed (2 icons).  

\begin{figure*}
    \centering
    \includegraphics[width=12cm]{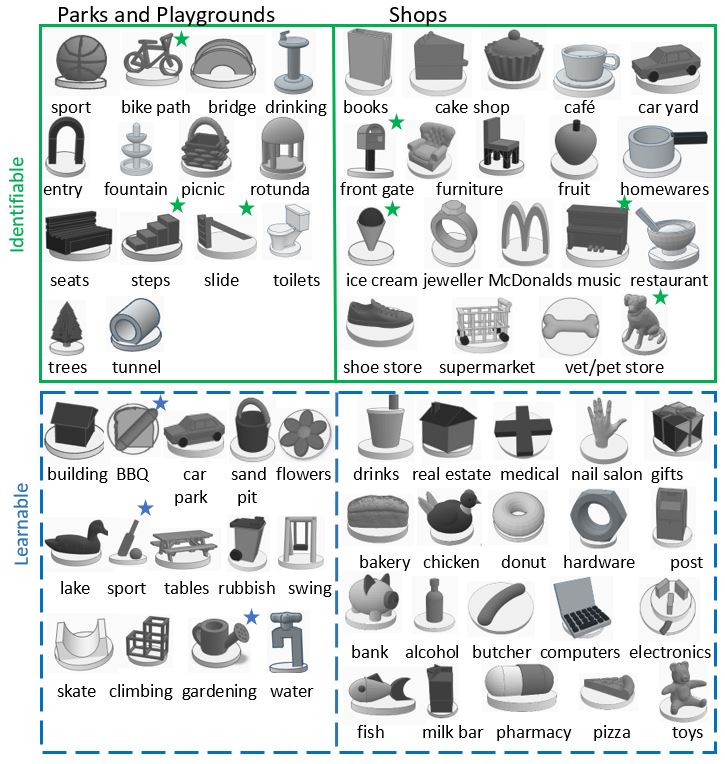}
    \caption{The selected TactIcons for use on 3D tactile maps of parks, playgrounds and shops. A star indicates those TactIcons that were introduced in the Follow-up Study.}
    \Description{Identifiable icons for parks and playgrounds: ball (sport), bicycle* (bike path), arched bridge (bridge), drinking fountain, archway (entry), 3-tiered fountain, picnic basket (picnic), rotunda, bench seat (seats), 4 steps* (steps), slide*, toilet with cistern (toilets), pine tree (trees), tunnel. 
    Identifiable icons for shops: book, slice of cake with cream filling and cupcake with patty pan and cherry on top (cake shop),  cup and saucer (cafe),  sedan car (car yard), letter box* (front gate), armchair and dining chair (furniture), apple (fruit), saucepan (homewares), round icecream scoop in pointed cone* (icecream), ring with jewel (jeweller), McDonalds M (McDonalds), upright piano* (music), bowl with chopsticks (restaurant), loafer (shoe store), supermarket trolley (supermarket) dog bone and Labrador dog* (vet/pet store). 
    Learnable icons for parks and playgrounds: building with roof (building), sausage on bread* (BBQ), sedan car (car park), bucket with handle up (sand pit), daisy head (flowers), duck (lake), cricket bat with ball* (sport), picnic table with seats (tables), wheelie bin (rubbish), swing in a-frame (swing), half pipe (skate), cubic climbing frame (climbing), watering can* (gardening), tap (water). 
    Learnable icons for shops: disposable cup with lid and straw (drinks), building with roof and chimney (real estate), flat cross (medical), outstretched hand (nail salon), package tied with string and bow on top (gifts), loaf of bread (bakery), chicken, donut, nut (hardware), post box (post), piggy bank (bank), wine bottle (alcohol), sausage (butcher), laptop (computers), power plug (electronics), upright fish (fish), open milk carton (milk bar), capsule (pharmacy), pizza slice (pizza), teddy bear (toys).}
    \label{fig:TactIcons}
\end{figure*}

The results from Study 2 were combined with those from Study 1 to identify a total of 33 3D icons that are easily recognisable, with 80\% or better recognition by both blind and sighted touch readers. These icons are listed in Table~\ref{tab:recognisable} and illustrated in Figure~\ref{fig:TactIcons}. 

\subsubsection{Learnable Icons}
Given the finding in Study 1 that 85\% of icons not identified on the first attempt could be remembered and recognised on a second attempt, it seems reasonable to expand the list of useful 3D tactile icons to a second tier. We define these as `learnable icons' -- for some touch readers they will require the initial use of a legend but they are then identifiable and are distinguishable from other icons in the corpus.

The results from Studies 1 and 2 reveal 34 learnable icons. 
Seven of these icons require a legend purely to define their association.
Six icons require a legend due to medium identification by novice touch readers. 
Five icons require a legend due to medium identification by congenitally blind touch readers. 
A further seven icons require a legend due to medium identification by both congenitally blind and novice touch readers.
These `learnable' icons are listed in Table~\ref{tab:learnable} and illustrated in Figure~\ref{fig:TactIcons}.
\subsection{Methodology for Touch Testing 3D Printed Tactile Icons}
Testing with both novice touch readers and congenitally blind people proved valuable in several instances. A Labrador dog was preferred to a cartoon dog by all of the blind participants (including the blind novice touch reader), with comments relating to their familiarity with Labrador guide dogs. For sports, the cricket bat was preferred by the congenitally blind participants over cricket stumps, which are more visually recognisable, because they were familiar with holding the bat when playing blind cricket.  Finally, we noted in Study 1 that blind (but not sighted) participants suggested that some icons be laid flat. This was tested with three icons in Study 2. Barbeque tongs and a butcher's knife were preferred in their original upright position, which enables access from all sides. All three congenitally blind touch testers preferred a fish to be laying flat because they could feel the shape of the tail more easily, but the novice touch testers  struggled to perceive a flat shape and preferred it upright.

\section{Discussion}
\subsection{Research Challenge 1: Proving the Potential Value of 3D Tactile Icons}
While decades of effort have been dedicated to researching tactile perception of point symbols for 2.5D tactile maps~\cite{Rener1993tactile,Rowell2003Taxonomy}, research into 3D symbols is in its infancy. Building on an in-the-wild study with only nine icons~\cite{Holloway20193D}, the current work is the first controlled study of representational 3D tactile icons. A total of 33 icons were found to be recognisable without the aid of a legend by 80\% or more of both congenitally blind and novice tactile readers. This success rate exceeds the ISO requirement of 66\% recognition for print icons, which is often not achieved~\cite{Davies1998, Forsythe2011}. In addition, 34 learnable icons were developed, which once deciphered do not need further consultation with a legend.
This compares favourably with 2.5D tactile icons, which are generally abstracted, require a legend for interpretation, and are limited to a maximum of 10-15 symbols per map~\cite{Lobben2012,Rowell2003world}.
It seems likely that the use of representational icons, rather than more abstract symbols, will reduce cognitive load for tactile map readers, be more inclusive and also be more engaging. 

This large corpus of easily identifiable icons opens the possibility for standardisation of tactile symbols. Unlike abstract raised line point symbols, the variety of icons means that the same 3D icons  can be used across multiple maps to convey the same map features. Furthermore, 3D printing enables the icons to be readily shared and reproduced. 

\textbf{Given the utility of these representational 3D printed tactile icons we feel they deserve a name -- TactIcons}, a tactile equivalent to visual icons and audio earcons~\cite{Blattner1989earcons}.

\subsection{Research Challenge 2: General Principles for the Design of 3D Tactile Icons}
Addressing Research Challenge 2, findings from the three studies enabled the compilation of a richer set of general principles for the design of 3D tactile icons.
\begin{enumerate}
    \item Icons must be tactually distinct:
        \begin{enumerate}
        \item Use objects that are a distinctive shape. Objects that are regular shapes such as rectangular prisms and rounded objects cannot easily be distinguished unless they have very distinctive additional features. For example, a round pizza, domed bicycle bell and rectangular barbeque all had 0\% recognition rates. 
        \item The most important and distinctive features should be near the top of the icon, where they can be more easily accessed by touch~\cite{Holloway20193D}. It may be necessary to position the object so that the distinctive features can be felt. For example, a coat hanger was easier to understand when laid flat because the  hook was difficult to feel when it was positioned upright.
        \item Simplify the object to its bare essentials~\cite{Google2021}, removing any extraneous details. It should be noted that the ideal level of detail may differ according to the user's tactile acuity. For example, some touch readers were able to distinguish tightly spaced legends on a model laptop and suggested making the keyboard layout more accurate, whereas others could only detect a rough texture. 
        \item Exaggerate features that are distinctive. For example, a dog or cat's ears should be exaggerated to assist with identification. 
        However, this exaggeration of features should not be stylistic, as we found that graphic shapes from popular culture such as a slice of bread with a stylistic top were much less easily recognised by congenitally blind people as they were unfamiliar with visual conventions.  
        \item Flatter icons are more difficult for novice touch readers to recognise than those that are more 3-dimensional in nature. 
        \item Corners and edges should be square or round in accordance with the original object to assist with interpretation. Often this means square corners for man-made objects and rounded corners for organic objects. This is in contrast with visual icon design guidelines that recommend consistency of corners~\cite{Google2021}. 
        \item Raised lines and parts are much easier to feel than holes or indents. If an indentation is important, consider outlining it with a raised line. 
        \item Larger icons are easier to feel. We tested our icons at a small scale (no more than 20mm\textsuperscript{3}), however they can and should be enlarged if space allows. 
        \item Icons must be spaced apart to allow adequate access for tactile exploration from all sides~\cite{Holloway20193D}. 
    \end{enumerate}
\item Icons should be based on concepts that are familiar to people who are congenitally blind~\cite{DAngiulli2007raised}:
    \begin{enumerate}
        \item Most importantly, icons should be co-designed with early input from people who are blind. 
        \item Objects that can be hand held are more familiar to people who are congenitally blind than those that are microscopic, too large to reach, rare, dangerous or intangible.
        \item Do not use visual icons unless they are widely discussed in popular culture. The McDonalds icon was only  recognised successfully because it is commonly described as `golden arches'. 
        \item Position icons in their most familiar orientation. For example the McDonald's logo was recognised more easily when standing upright. However, note that orientation of an icon should also be determined by touch access, as per 1(c).   
    \end{enumerate}
\item Support decoding with context and accompanying information: 
    \begin{enumerate}
        \item Provide a map title, as knowing the map type reduces the possibilities and assists with association between the object and its meaning~\cite{Heller1996tactual,Rogers1989icons}.
        \item If possible, represent objects that are all around the same size (hand-held is best) to avoid confusions of scale.
        \item Provide a choice of media, for example incorporate clear print, braille and audio labels along with the TactIcons. 
    \end{enumerate}
\item Consider the context and user. It is unlikely that a single set of TactIcons can be universally understood. Instead, a range of options may be needed: 
\begin{enumerate}
    \item Some icons may need to differ according to culture. For example, power plugs differ across regions. 
    \item Consider the expected age of the map users. Some icons differ in their recognition according to generation, for example the typical shape of rubbish bins has changed throughout the years. 
\end{enumerate}

\end{enumerate}
\subsection{Research Challenge 3: A Practical Methodology to Co-Design and Test the Efficacy of 3D Tactile Icons}
We trialed our proposed methodology for evaluating 3D tactile icons (Figure \ref{fig:testing_method}) in Study 2. A subset of 40 new or revised TactIcons were presented to both novice touch readers (sighted, vision impaired and blind) to ensure the icons were tactually distinct, along with people who are congenitally blind to ensure the icons were based on concepts that were familiar to them. 

This proposed methodology was successful, allowing us to further refine our set of recommended TactIcons. As it was sometimes difficult to make a decision about the effectiveness of an icon due to the wide variation between individuals, we would not recommend using less than three novice and three congenitally blind touch testers. More testers would obviously be advantageous, however we acknowledge that this may not always be practical -- evaluation is labour-intensive and accessible formats teams have greatly reduced their workforce of touch readers over the past two decades. The testing methodology aligns well with best practice guidelines for the design and verification of print icons~\cite{buhler2022UIPP,Nolan1989}.   

It should be noted that we used sighted participants for the first round of touch testing as a matter of convenience, as they are a more readily available participant pool within our workplace. The opposite may actually be true in other contexts where tactile maps are developed, such as accessible formats production teams with dedicated touch proof readers, or when developing maps of workplaces or schools used by a number of blind staff or students. In these cases, novice touch readers (either sighted or with recent vision loss) will still be required to confirm that the materials are tactually distinct, but could do so after pre-screening by BLV testers.

Only a minority of the co-design suggestions for refinements and new icons led to improvements, as illustrated in Figure~\ref{fig:BBQ}. It was helpful to test all of the co-design suggestions for the purposes of this research, however for practical implementation we advise first cross-referencing ideas against the design guidelines.  

\subsection{Developing a Verified Set of 3D Icons}
Using both the design guidelines and recommended methodology for testing 3D icons, this project developed, evaluated and selected 33 icons that are reliably recognisable and a further 34 icons that are learnable. These TactIcons have been shared at \url{https://www.thingiverse.com/thing:5841775}.
The icons were developed for use on maps of parks, playgrounds and shops but some could also be applied to other contexts. For example, the icons for toilet, rubbish bin, entrance and restaurant can be applied to indoor and event maps. 

A successful icon was not identified for 13 of the 84 map features explored. Further iterative co-design may result in successful icons for some of these features, but it is likely that some map features will still need to be represented using abstract symbols and/or in conjunction with a legend.

\subsection{Limitations and Future Work}
The design of map symbols is an art rather than a science. As stated by MacEachren, ``there is no single correct scientific, or nonscientific, approach to how maps work''~\cite{MacEachren2004}. Different designers and testers will inevitably devise different icons for the same map features, and the 3D tactile icons that we have recommended for use are not necessarily the best and only possible designs. In particular, different icons may be needed depending on region and age of the intended users. Nevertheless, our work offers a pragmatic methodology for ensuring the development of TactIcons that are effective within the context that they are designed and can be used by people who are BLV with a wide range of skill sets. 

Due to individual differences in mental imagery, TactIcons may not be helpful for some users, and should therefore be combined with other options such as clear print, braille and audio labels.

An important next step will be the evaluation of TactIcons on maps, both as part of laboratory studies compared with abstract tactile symbols, and in-the-wild studies where the icons are used on real world tactile maps.  The value of TactIcons may be impacted by proximity to other tactile elements, use in combination with a legend and other labelling methods, and use for more complex cognitive tasks. Quantitative measures such as the time taken to correctly guess an icon or the number of incorrect guesses were not recorded as part of this paper due to practical restrictions with a large number of icons being tested at once. However, this may be a useful final step in future work to confirm the appropriateness of icons once they have been refined and selected.

While 3D printing was leveraged in this work because it is well-suited to rapid prototyping and is being adopted by accessible format providers, TactIcons can of course be produced by signage companies using other manufacturing processes more suitable for permanent and outdoor installations. 

The icons in this work were designed and tested purely on the basis of their use for touch recognition. However, there is a strong argument for the creation of tactile maps for engagement with all users~\cite{Jansen2015opportunities,Kent2019maps}. A broadening of tactile maps for universal access would also bring them to a more prominent position, wider adoption and ultimately provide a more inclusive experience for touch readers. TactIcons therefore need to function for both tactile and visual use. Given that the majority of people with a vision impairment are not totally blind, but instead have some level of vision~\cite{Bourne2021}, the addition of  meaningful colours and high contrast should also be considered in the design of TactIcons.

\section{Conclusion}

This work provides three major contributions. First, it verifies the utility of 3D tactile icons, or TactIcons, as an intuitive and easy-to-understand means of indicating features on tactile maps. Whereas raised line maps rely on a maximum of 15 abstract point symbols that must be interpreted using a legend, a much larger number of TactIcons can be distinguished from one another and can be understood without reference to a legend, potentially reducing the significant cognitive demands of interpreting tactile maps. 

Most importantly, this paper proposes design guidelines for TactIcons along with a practical methodology for testing these icons, based on empirical findings regarding performance differences between blind and sighted touch testers.  
Newly designed tactile icons must be touch tested by both novice touch readers to check that the icons are tactually distinct and by people who are congenitally blind to ensure icons are based on concepts with which they are likely to be familiar.  

The third and final contribution of this paper was utilisation of the proposed design principles and testing methodology to create a set of 67 3D icons for use on tactile maps of parks, playgrounds and retail shopping areas. The icons have been shared at \url{https://www.thingiverse.com/thing:5841775} for use by tactile map designers to assist in the creation of inclusive tactile maps. 

It is hoped that this work will assist with the creation of user-friendly tactile maps for people who are blind or have low vision, thereby supporting independent travel and educational success. The use of representational 3D tactile icons on maps has the added advantage of being useful and appealing to the general public, creating an inclusive experience for all. 
\begin{acks}
This work was conducted as part of a project investigating the use of 3D printing for touch readers, supported by the Australian Research Council LP170100026.  Thanks are extended to our ARC Linkage Project Partners: the Round Table on Information Access for People with Print Disabilities Inc., the Department of Education Victoria, Guide Dogs Victoria, NextSense and the Royal Society for the Blind. Thanks also to Kate Stephens for her input into the initial TactIcon designs and to the many touch testers for their participation and insights.  
\end{acks}
\bibliographystyle{ACM-Reference-Format}
\bibliography{3Dsymbols}
\clearpage
\onecolumn
\appendix
\section{Appendix: Participant Information}
\begin{table*}[!hb]
    \centering
      \caption{Participant details across Studies 1 and 2. Results were pooled for participants at public events, listed in the first row. }
    \begin{tabular}{llrrrr}
    \toprule
   \thead{Vision Level} & \thead{Touch Reading\\Experience} & \thead{Age} & \thead{Study 1\\Icons (n)} & \thead{Study 2\\Icons (n)} & \thead{Overall\\Accuracy}\\
   \midrule
 14 sighted \& 1 low vision & novices & 18-70; $\bar{x}=42$ & 117& 0 & 49\%\\
   sighted & novice & 13& 182& 0&68\%\\
   sighted & novice & 15& 156& 0&50\%\\
   sighted & novice & 20& 15& 0&27\%\\
   sighted & novice & 25& 5& 0&60\%\\
   sighted & novice & 31 & 20 & 0 & 50\%\\
   sighted & novice & 37 & 50 & 0 & 60\%\\
   sighted & novice & 38 & 124 & 0 & 51\%\\
   sighted & novice & 39 & 54& 0 & 39\%\\
   sighted & novice & 41 & 30 & 0 & 53\%\\
   sighted & novice & 42 & 4 & 0 & 50\%\\
   sighted & novice & 43 & 28 & 0 & 57\%\\
      sighted & novice & 46 & 85& 0 & 76\%\\
      sighted & novice & 49 & 169 & 43 & 59\%\\
   sighted & novice & 52 & 37 & 0 & 59\%\\
   sighted & novice & -- & 29& 0 & 65\%\\
   sighted & novice & -- & 59& 0 & 90\%\\
   sighted & novice & -- & 67 & 0 & 59\%\\   
     low vision & novice & 63 & 70& 49& 65\%\\
   blind & novice & 60 & 0 & 61 & 34\%\\
   blind & intermediate & 44 & 85 & 0 & 65\%\\
   blind & intermediate & 67 & 87 & 0 & 55\%\\
   blind & expert & 24 & 86 & 0 &  79\%\\
   blind & expert & 28  & 86 & 0 &  79\%\\
   blind & expert & 29 & 87 & 52 &  75\%\\
   blind & expert & 39 & 85 & 0 &  78\%\\
   blind & expert & 51 & 86 & 0 &  72\%\\
   blind & expert & 57 & 85 & 0 &  71\%\\
   blind & expert & 57 & 88 & 41 & 83\%\\
   blind & expert & 61 & 0 & 56 &  66\%\\
   blind & expert & 64 & 87 & 0 &  21\%\\
   blind & expert & 65 & 132 & 0 & 70\%\\
\bottomrule
\end{tabular}
\label{tab:participants}
\end{table*}

\clearpage
\section{Appendix: Recognisable TactIcons}
\begin{table*}[h!]
    \centering
    \caption{Recognisable 3D tactile icons for which the object could be recognised and associated with the intended map feature by 80\% or more of all touch testers in Study 2 and the Follow-up Study. Icons introduced in the Follow-up Study are marked with an asterisk*.}
        \begin{tabular}{l|rrr|rrrr}
    \toprule
     & & Sighted & & & Blind & & \\ 
   \thead{Icon} & \thead{Recognition\\(\%)} & \thead{Association\\(\%)} & \thead{\\ n} & \thead{Recognition\\(\%)} & \thead{Association\\(\%)} & \thead{Recall\\(\%)} & \thead{\\n} \\
   \midrule
   \textbf{Parks and Playgrounds}& & & & & & & \\
    Ball games -- ball  & 100 & 100 & 10 & 82 & 100& 100& 11\\
    Bicycle path -- bicycle & 90 & 100 & 10 & 91 & 100& 100& 11\\
    Bridge -- arch bridge  & 90  &89 &10 & 91 & 100&100 & 11\\
    Drinking fountain & 90 & 100& 10 & 82 & 100& 100& 11\\
    Entry -- archway & 80 & 50 & 10 & 91 & 100& --&11 \\
    Fountain & 80 & 100 & 10 & 82& 100& 100& 11\\
    Picnic area -- picnic basket & 80 & 100 & 10 & 100 & 82& n/a&11\\
    Rotunda & 80 & 88 &10 & 80 &100&--&10\\
    Seats -- park bench & 100&100 &10  & 100 & 91 & n/a& 11\\
    Steps -- 4 steps & 80 & 88 & 10 & 100 & 100& n/a      & 11\\
     Slide -- straight slide & 90&100 & 10 & 82 & 100& 100& 11\\
    Toilets -- toilet & 80 & 100 & 10 & 91 & 100& --& 11\\
    Trees -- pine tree & 100 & 100 & 10 & 82 & 100&100 & 11\\
    Tunnel & 100 & 100 & 10 & 90 & 100& 0& 10\\
    \midrule
       \textbf{Shops}& & & & & & & \\
         Bookstore or library -- book & 90& 100&10 & 82& 100& 100& 11\\
    Cake shop -- cake slice & 90 & 100 & 10 & 82 & 100&100& 11\\
     Cake shop -- cupcake with cherry & 82 & 100  & 11 & 82& 100& 100& 11\\
     Car yard -- sedan car & 82 & 100 & 11 & 91 & 90 & 100 & 11\\
    Coffee shop -- teacup & 100 & 100 & 11 & 91 & 100& 100& 11\\
    Front gate -- letter box* & 100\% & n/a & 3 & 100\% & n/a & n/a& 3\\
        Furniture store -- armchair & 100 & 80 & 10 & 100& 80& n/a& 10\\
        Furniture store -- dining chair* & 100 & 83 & 7 & 100 & 100 & n/a & 5\\ 
    Greengrocer -- apple & 90 & 100& 10& 82& 100& 100& 11\\
    Homewares/kitchen -- saucepan & 80 & 88 & 10 & 91 & 100& --& 11\\
    Ice cream -- round scoop in cone* & 100 & 100 & 2 & 100 & 100 & n/a & 4\\
      Jewellery -- ring & 100 & 100 & 10 & 100 & 100& n/a& 10\\
 McDonalds -- upright arches & 90 & 100 & 10 & 100 & 90 & n/a& 11\\
 Music -- upright piano & 100 & 100 & 3 & 100 & 100 & n/a & 3\\
 Restaurant -- bowl with chopsticks & 100& 100& 10 & 90 & 100& 100& 10\\
 Shoe store -- loafer & 90 & 100  & 10 & 80 & 100& 100&10 \\
     Supermarket -- trolley & 100& 100& 10& 91 & 100& --& 11\\
  Vet or pet store -- bone & 100& 100& 10& 90 & 100& --& 10\\
  Vet or pet store -- Labrador* & 100 & 100 & 7 & 100 & 100 & n/a & 5\\
    \bottomrule 
    \end{tabular}
    \label{tab:recognisable}
\end{table*}
\clearpage
\section{Appendix: Learnable TactIcons}
\begin{table*}[h!]
    \centering
      \caption{Learnable 3D tactile icons recommended for use but where a legend is required for initial confirmation because of inconsistencies in Association (knowing what map feature the object represents) or Recognition (identifying the object) by sighted or blind touch testers in Study 2 and the Follow-up Study. Icons introduced in the Follow-up Study are marked with an asterisk*.}
    \begin{tabular}{l|rrr|rrrr}
    \toprule
     & & Sighted & & & Blind & & \\ 
   \thead{Icon} & \thead{Recognition\\(\%)} & \thead{Association\\(\%)} & \thead{\\ n} & \thead{Recognition\\(\%)} & \thead{Association\\(\%)} & \thead{Recall\\(\%)} & \thead{\\n} \\
   \midrule
   \textbf{Association} & & & & & & & \\
    Building -- cube with roof & 80 & 50 & 10 & 100 & 64& n/a& 11\\
    Car park -- sedan & 80 & 100 & 10 & 82&78&100&11\\
     Drinks -- cup with straw & 90 & 89 & 10 &90& 67 & n/a& 10\\
     House -- building with chimney & 90 & n/a &10 & 91 & 40& 100 &  11\\
     Medical -- flat cross & 100&90 &10  & 82 & 63& 100& 11\\
      Nail salon -- hand & 90 & 56 & 10 & 90 & 44&100 &10 \\
    Sand pit -- bucket with handle up & 90 & 75 & 10 & 90 & 67& --& 10\\
\midrule
\textbf{Recognition by novices}& & & & & & & \\
Garden bed -- daisy head & 70 & 100 & 10 & 91 & 100& --& 11\\
Gift store -- box with bow & 50& 100& 10 & 91 & 89& 100& 11\\
Lake -- duck & 50 & 83 & 12 & 82 & 100& 100&11\\
Oval or sports -- cricket bat \& ball* & 66 & 100 & 3 & 100 & 100 & n/a & 3 \\
Picnic table & 50 & 100 & 10 & 82&100 & 0&11 \\
Rubbish -- wheelie bin & 70& 100& 10 & 80 & 100& 100&10 \\
Swing & 50 &100 & 10& 80 & 100& -- & 10\\
\midrule
\textbf{Recognition by blind}& & & & & & & \\
 Bakery -- bread loaf & 80& 100& 10& 55& 100 &80 & 11\\
    Skate park -- half pipe & 100 &100 & 11 & 45 &80&100&11\\
Climbing -- square climbing frame & 100 & 83 & 10 & 64 &100&100&11\\
 Chicken shop -- chicken sitting & 100 & 90 & 10 & 64& 100& 100 & 11\\
 Community garden -- watering can* & 100 & 100 & 5 & 75 & 100 & -- & 4\\
  Donuts -- donut &80 &100 &10  & 73 & 100 & 100& 11\\
 Hardware -- nut & 90 & 100 & 10 & 45 & 75& 100 & 11\\
  Post office -- mail box & 80 & 100 & 10 & 64 & 100 & 100& 11\\
\midrule
\textbf{Recognition by all}& & & & & & & \\
Bank -- piggy bank & 67 & 63 & 12 & 64 & 83& 100& 11\\
BBQ -- sausage on bread* & 71 & 100 & 7 & 50 & 100 & -- & 4\\
Bottle shop, bar or pub -- wine bottle & 60 & 100& 10& 73& 88& 67& 10\\
Butcher -- sausage & 60 & 100 & 10 & 50& 100& 100& 10\\
Computers -- laptop & 50 & 100 & 10 & 55& 100& 100& 11\\
Electronics -- power plug & 60 & 100 & 10 & 73& 91& 100& 11\\
Fishmonger or aquarium -- upright fish& 70 & 86 & 10 & 55& 100& 100& 11\\
Florist -- sunflower in vase & 60& 100& 10& 55& 100& 75& 11\\
 Milk bar -- open milk carton & 70 & 100 & 10 & 70 & 86& 100& 10\\
 Pharmacy -- capsule & 78 & 100 & 9 & 67& 100& 100& 9\\
  Pizza -- pizza slice & 60& 100& 10 & 64 & 100& 100& 11\\
 Toy store -- teddy bear & 70 & 100 & 10 & 73 &100 & 50& 10\\
Water -- tap & 70 & 100 & 10 & 60 & 100& 67& 10\\
\bottomrule
    \end{tabular}
    \label{tab:learnable}
\end{table*}

\end{document}